\documentclass[12pt,preprint]{aastex}
\newcommand{\simle}{\mbox{$\stackrel{<}{_{\sim}}$}}

\newcommand\lsun{\hbox{\,L$_\odot$}}

\shorttitle{Size-Luminosity Relations}
\shortauthors{Monnier et al.}

\begin{document}




\title{The near-infrared size-luminosity relations for Herbig Ae/Be disks}


%% author and affiliation information
%% Note that \email has replaced the old \authoremail comman
%% from AASTeX v4.0. You can use \email to mark an email addres
%% anywhere in the paper, not just in the front matter
%% As in the title, you can use \\ to force line breaks

\author{J.~D.~Monnier\altaffilmark{1},
R.~Millan-Gabet\altaffilmark{2}, R.~Billmeier\altaffilmark{1}, R.~L.~Akeson\altaffilmark{2},
D.~Wallace\altaffilmark{3},
J.-P.~Berger\altaffilmark{4}, 
N.~Calvet\altaffilmark{5}, 
P.~D'Alessio\altaffilmark{6},
W.~C.~Danchi\altaffilmark{3},
L.~Hartmann\altaffilmark{5}, L.~A.~Hillenbrand\altaffilmark{7}, 
M.~Kuchner\altaffilmark{8}, J.~Rajagopal\altaffilmark{3},
W.~A.~Traub\altaffilmark{5},
P.~G.~Tuthill\altaffilmark{9}, A.~Boden\altaffilmark{2},
A.~Booth\altaffilmark{10}, M.~Colavita\altaffilmark{10},  J.~Gathright\altaffilmark{11}, 
M.~Hrynevych\altaffilmark{11}, D.~Le~Mignant\altaffilmark{11},
R.~Ligon\altaffilmark{10}, C.~Neyman\altaffilmark{11}, M.~Swain\altaffilmark{10},
R.~Thompson\altaffilmark{2},
G.~Vasisht\altaffilmark{10}, P.~Wizinowich\altaffilmark{11}, C.~Beichman\altaffilmark{2},
J.~Beletic\altaffilmark{11},
 M.~Creech-Eakman\altaffilmark{10}, C.~Koresko\altaffilmark{2}, 
 A.~Sargent\altaffilmark{2},
M.~Shao\altaffilmark{10}, 
\& G.~van~Belle\altaffilmark{2}
}

\altaffiltext{1}{monnier@umich.edu: University of Michigan Astronomy Department, 
941 Dennison Bldg, Ann Arbor, MI 48109-1090, USA.}
\altaffiltext{2}{Michelson Science Center, California Institute of Technology, 770 South Wilson Avenue, Pasadena, CA 91125}
\altaffiltext{3}{NASA Goddard Space Flight Center}
\altaffiltext{4}{Laboratoire d'Astrophysique de Grenoble, 414 Rue de la Piscine 38400 Saint Martin d'Heres, France}
\altaffiltext{5}{Harvard-Smithsonian Center for Astrophysics, 60 Garden St,
Cambridge, MA, 02138, USA}
\altaffiltext{6}{Universidad Nacional Aut\'{o}noma de M\'{e}xico}
\altaffiltext{7}{Astronomy Department, California Institute of Technology, Pasadena, CA}
\altaffiltext{8}{Princeton University, Princeton, NJ}
\altaffiltext{9}{University of Sydney, Physics Department}
\altaffiltext{10}{Jet Propulsion Laboratory, California Institute of Technology,  4800 Oak Grove Drive, Pasadena, CA 91109}
\altaffiltext{11}{W. M. Keck Observatory, California Association for Research in Astronomy, 65-1120 Mamalahoa Highway, Kamuela, HI 96743}

%\email{Contact: monnier@umich.edu}


\begin{abstract}
We report the results of a sensitive K-band survey of Herbig Ae/Be
disk sizes using the 85-m baseline Keck Interferometer.  Targets were
chosen to span the maximum range of stellar properties to probe the
disk size dependence on luminosity and effective temperature.  For
most targets, the measured near-infrared sizes (ranging from 0.2 to
4~AU) support a simple disk model possessing a central optically-thin
(dust-free) cavity, ringed by hot dust emitting at the expected
sublimation temperatures ($T_s\sim$1000-1500K).  Furthermore, we find
a tight correlation of disk size with source luminosity $R\propto
L^{\frac{1}{2}}$ for Ae and late Be systems (valid over more than 2
decades in luminosity), confirming earlier suggestions based on
lower-quality data.  Interestingly, the inferred dust-free inner
cavities of the highest luminosity sources (Herbig B0-B3 stars) are {\em
under-sized} compared to predictions of the ``optically-thin cavity''
model, likely due to optically-thick gas within the inner AU.

\end{abstract}



%\tableofcontents

\section{Introduction}

Young stellar objects (YSOs) are often observed to be surrounded by an
optically-thick accretion disk, presumably left-over from early
stages of star formation.  These disks are expected to evolve into
optically-thin ``debris'' disks as the circumstellar material either
accretes onto the central star, is blown out of the system, or
coagulates into planetessimals.  The disappearance of the
optically-thick protostellar accretion disk thus marks a transition
between the final stages of star formation and the onset of planet
formation.  High angular resolution studies of these transition disks
effectively test our physical models of accretion as well as reveal
the initial conditions for planet-building.  In particular, infrared
interferometers directly probe the temperature and density structure
of gas and dust within the inner AU of YSO disks, critical ingredients
in any recipe for planet formation.

\citet{rmg1999a} first discovered the marked discrepancy between
theoretical predictions and the observed near-infrared (NIR) sizes of
Herbig Ae/Be stars using the Infrared-Optical Telescope Array
(IOTA) Interferometer -- the AB~Aur disk
was found to be many times larger than expected.\footnote{In this
paper, the ``disk size'' generally refers to the extent of the disk
emission at a particular wavelength, not the {\em physical} extent of
all the disk material.}  This pattern has been confirmed again and
again, with ``large'' disk sizes for Herbig Ae/Be stars being found by
Keck aperture masking \citep{tuthill2001a,danchi2001}, additional IOTA
work \citep{rmg2001}, and Palomar Testbed Interferometer observations
\citep{akeson2000,akeson2002, eisner2003,eisner2004}.  More recently,
a similar pattern has been found for bright T~Tauri disks too
\citep{akeson2005,akeson2000,colavita2003}.

Popular disk theories of that time
\citep[e.g.,][]{l-bp1974,adams1987,calvet1991,hillenbrand1992,hartmann1993,cg97}
incorporated an optically-thick, geometrically-thin disk (with
flaring, which is not so relevant for NIR wavelengths).  In these
models, the optically-thick disk midplane shields inner dust from much
of the stellar radiation, thus hot dust near sublimation temperature
($T_s\sim1500K$) can exist quite close to the star.  Interferometer
measurements showed this emission to be much further from the stars
than these models predicted, thus arose the size controversy.  About
the same time, \citet{natta2001} revisited the problem of the
unrealistically high accretion rates inferred for some Herbig Ae/Be
stars \citep[based on NIR excess;][]{hillenbrand1992,hartmann1993} 
using spectroscopic
observations from the Infrared Space Observatory (ISO).  These two
observational ``problems'' would be resolved by the same solution.

First suggested by \citet{tuthill2001a} and independently developed by
\citet{natta2001} and \citet{dullemond2001}, the large disk sizes and
excess NIR flux were naturally explained by the presence of an {\em
optically-thin} cavity surrounding the star.  Thus, the innermost disk
is {\em not} optically thick, presumably because dust, which is the
primary source of opacity for $T\simle1500K$, is absent due to
evaporation by the stellar radiation field.  This geometry explains
the NIR excess flux as well since a frontally-illuminated dust wall
efficiently emits in the NIR.  Interestingly, earlier modellers
had already realized the inner cavity would be devoid of dust due to
the high temperatures \citep[e.g.,][]{hillenbrand1992}, but the
temperature profile adopted by these workers still implicitly
incorporated an optically-thick disk midplane.  These central cavities
are not necessarily devoid of gas; indeed, the optical depth of the
gas depends on many factors (most notably, the accretion rate and
geometry).  Central clearings in YSO disks are not only interesting in
the context of accretion disk physics but have been implicated in
halting migration of the extrasolar hot Jupiter planets
\citep{kuchner2002}.

\citet{monnier2002a} put the ``optically-thin cavity'' model to the
most stringent test thus far by analyzing the full set of published
interferometer measurements (including both T~Tauri and Herbig Ae/Be
stars for the first time) and found overall consistency through the
use of a ``size-luminosity'' diagram.  This diagram is particularly
powerful because the inner radius of dust destruction is nearly
independent of stellar temperature and almost a pure function of
luminosity.  This work also uncovered evidence for possible absorption
by the {\em gaseous} inner disk of the most luminous sources in the
sample, although interpretation was limited by significant data
scatter due (at least in part) to the heterogeneous nature of the
datasets.

Recently, \citet{eisner2003,eisner2004} made significant contributions
to the studies of these disks in a number of ways.  Firstly, they
measured elongated NIR emission from some Herbigs, evidence for
disk-like structure that had eluded the shorter-baseline IOTA work.
Furthermore, the size-luminosity relations of \citet{monnier2002a}
were confirmed for the Herbig Ae stars. Perhaps most interestingly,
\citet{eisner2004} presented the clearest evidence to-date for {\em
under-sized} disk emission around the early B stars in their sample.  In fact,
the NIR sizes of these disks were found consistent with the ``classical''
optically-thick, geometrically-thin disk models.

Within this context, our group has carried out a survey of YSOs
 as part of ``shared-risk'' commissioning of the
Keck Interferometer.  The Keck Interferometer boasts 10$-$100$\times$
the sensitivity over previous-generation instruments, allowing a large number of
Herbig Ae/Be, T Tauris, and FU~Orionis objects to be targeted.
In order to extend beyond existing work, we designed our survey as
follows.  First, great care was taken to choose targets with reliable
spectral types and luminosities (early interferometers could detect
relatively few targets and many had ambiguous classifications).
Second, the greater instrumental sensitivity allowed us to probe
systems spanning a larger range of spectral types and stellar
luminosities.

Here we report our results for the Herbig Ae/Be portion of the survey,
where we have significantly reduced the observational ``scatter'' that
hampered previous studies.  With the improved data quality, we can
definitively characterize the size-luminosity relations of Herbig
Ae/Be disks.

\section{Observations}

The Keck Interferometer (KI) was used during its visibility science
commissioning period (2002-2004) to observe 14~Herbig Ae/Be stars as
part of this survey (see Table~\ref{targets}).  The KI is formed
by two 10-m aperture telescopes (each consisting of 36 hexagonal mirror
segments) separated by 85~m along a direction $\sim$38~degrees East of
North, corresponding to a minimum fringe spacing of
5.3~milli-arcseconds at 2.2$\mu$m.  In order to coherently combine the
NIR light from such large apertures, each telescope utilizes a
natural guide star adaptive optics system \citep{wiz2003}.  Optical delay lines
correct for sidereal motion and the telescope beams are combined at a
beamsplitter before the light is focused onto single-mode (fluoride)
fibers which impose a $\sim$50~milliarcsecond (FWHM) field-of-view
for all data reported herein.  While both 
H and K-band 
observations are now possible, only broad K-band
(2.18$\mu$m, $\Delta\lambda= 0.3 \mu$m)
data are reported here.  Owing to the large apertures and excellent
site, the Keck Interferometer is currently the world's most sensitive
infrared interferometer, recently becoming the first such instrument
to measure fringes on an extra-galactic object \citep{swain2003}.
Further technical details  can be
found in recent Keck Interferometer publications 
\citep{colavita2003,cw2003,cw2000}.

Table~\ref{targets} summarizes the basic properties of the target
stars, including spectral type, distance, luminosity, and literature
references for photometry used herein.  Calibration of fringe data was
performed by interspersing target observations with those of
unresolved calibrators (see Table~\ref{calibrators}).  The square of
the fringe visibility (V$^2$) was measured using the ABCD-method 
\citep[using a dither mirror; see also][]{shao1977} and
we followed the well-tested strategies described for the Palomar
Testbed Interferometer \citep{colavita1999}, except that corrections
for uneven telescope ratios were improved and jitter corrections were
not applied.  We refer the reader to \citet{colavita2003} and
\citet{swain2003} for further description of calibration procedures.

The calibrated V$^2$ results appear in
Table~\ref{keckresults} along with with the projected baseline (u,v)
and date for each independent dataset.
The V$^2$ errors
reported in this table only include statistical errors.
Internal data quality checks have established a conservative upper limit 
to the systematic error $\Delta\,V^2=0.05$.  Model fitting in this paper
includes both sources of error in the uncertainty analysis.

\section{Methodology}
\label{methodology}
In this study, we wish to measure the characteristic NIR
sizes of YSO accretion disks.  By using a target sample spanning
a wide range of stellar properties, we aim also investigate the size
dependence on stellar luminosity.  Interferometers have been
successfully used to resolve stars since 1921 \citep{michelson1921},
and a full discussion of the methodology will not be given here.  A
single-baseline interferometer can effectively determine the
characteristic size of an astronomical object by measuring the fringe
``visibility,'' a measure of the fringe contrast; unresolved sources
produce high contrast fringes (visibility unity) while resolved
objects have low visibility \citep[for further discussion,
see][]{tms2001}.  Visibility data can be converted into a quantitative
``size'' estimate by using a model for the brightness distribution and
applying a Fourier Transform.  Since we can not {\em image} the disks
directly yet, we must adopt a simple empirical model capable of
parameterizing the spatial extent of the disk emission.

\citet{monnier2002a} discuss the merits of using a ``ring'' model for
describing the NIR emission from a circumstellar disk, based
on the argument that only the hottest dust at the inner edge of the
disk can contribute significantly to the NIR emission
\citep[see also][]{natta2001}.  Ring models have been fit to
visibility data by other workers in the field as well
\citep{millangabet2001,eisner2003,eisner2004}.  Furthermore, the first
imaging results for the LkH$\alpha$~101 disk \citep{tuthill2001a}
appear to support this class of models.

We have fitted our visibility data with a simple model consisting of a
point source (representing the unresolved stellar component) and a
thin (circular) ring with average diameter $\theta$ (representing the
circumstellar dust emission).  The thickness of the ring can not be
easily constrained for measurements on the main lobe of the visibility
curve (i.e, when the fringe spacing is less than the ring diameter)
and thus our results are only sensitive to the {\em average} ring
diameter, not the ring thickness.  For the fits reported here, we have
used a uniform brightness ring with a fractional thickness of 20\%,
inspired by the LkH$\alpha$~101 image.

Because our interferometer measurements lack the requisite
(u,v) coverage to fully constrain our ring model, 
the fraction of the total K-band emission coming from the disk must be
estimated separately through SED fitting.  
This method was first applied to interferometry observations 
of YSOs by \citet{rmg1999a} and is now in common usage
\citep{rmg2001,akeson2000,akeson2002,
eisner2003,eisner2004}.  Here, we fit the broadband SED with a two-component model, 
consisting of a stellar spectrum and a single-temperature dust blackbody.
We have slightly improved on the standard procedure by
using a Kurucz model \citep{kurucz1979} for the underlying stellar
spectrum instead of a simple blackbody, which makes some difference
for sources with small IR excess or cool stellar atmospheres.
Reddening of the central star must also be included in the fit, and here we
adopted the reddening law from
\citet{mathis1990}; this choice affects our (de-reddened) luminosity estimates
for the most obscured targets.

\begin{figure}[thb]
\begin{center}
%\epsscale{.5}
{
\includegraphics[angle=90,width=3in]{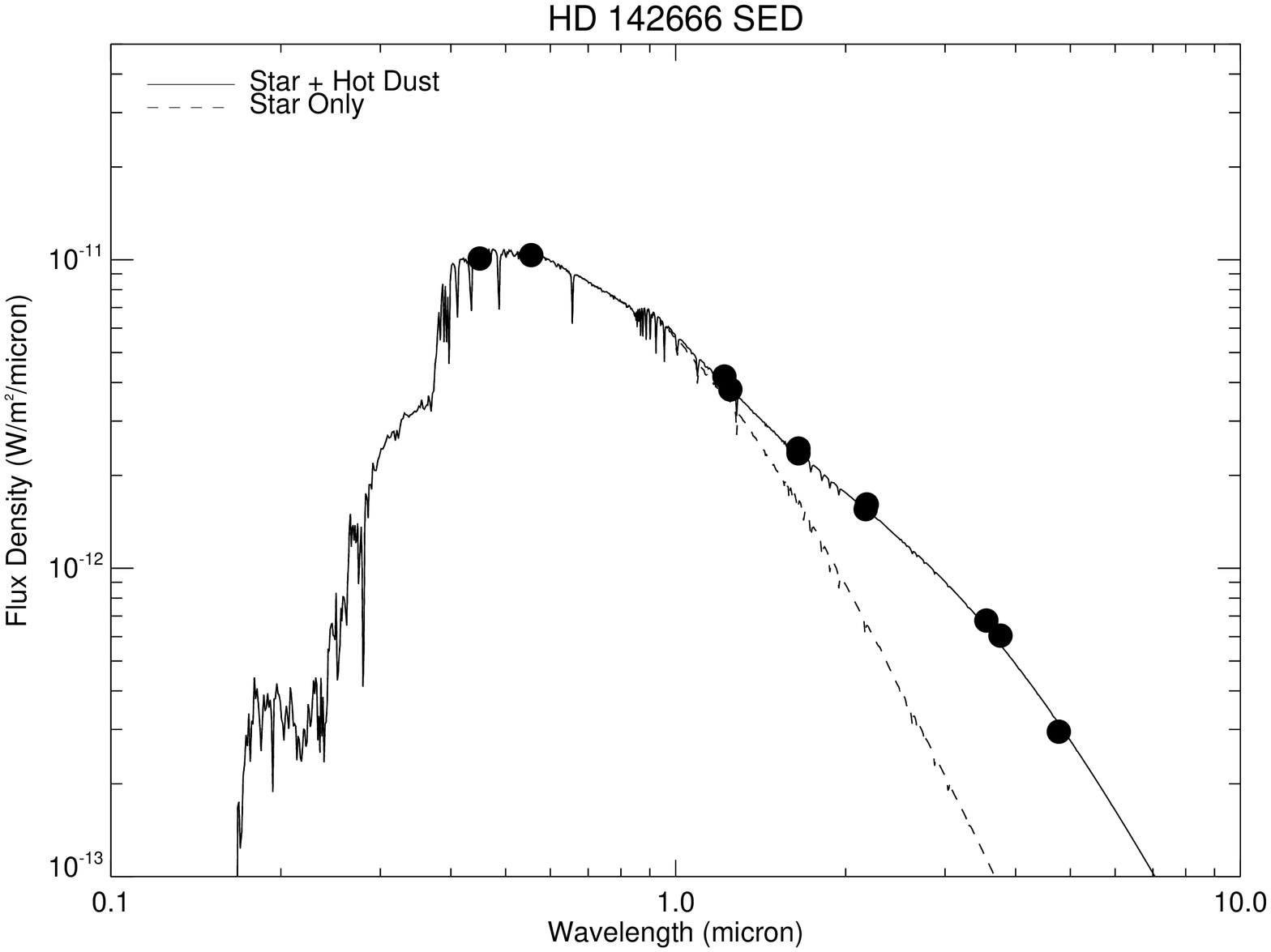}
\hphantom{.....}
\includegraphics[angle=90,width=3in]{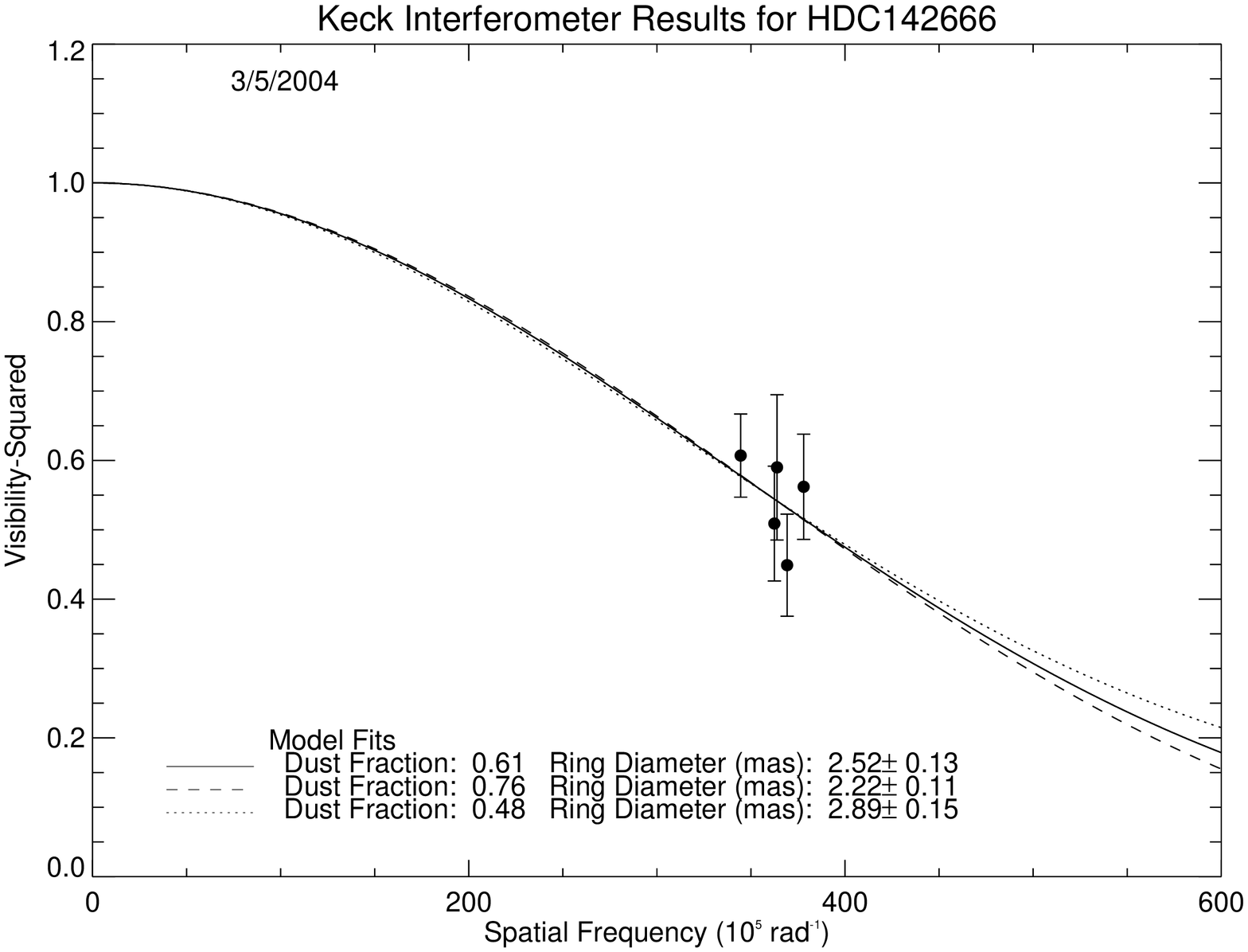}
}
%      \plotfiddle{Figure/fig:astrometry.eps,rot=90}^M
%\plotfiddle{Figures/fig:astrometry.eps}{12cm}{1}{1}{0}{0}{-90}^M
\caption{\scriptsize 
a) SED fit for HD~142666, including (reddened) A8V stellar
spectrum plus hot dust blackbody ($T=1355~K$).  Solid points are
photometry from the HST Guide Star Catalog \citep{morrison2001}, 2MASS
\citep{cutri2003}, and the Catalog of Infrared Observations
\citep{gezari1999}. b) Ring model fits to the Keck Interferometer
visibility data for three different estimates of the dust
fraction at 2.2$\mu$m. The three dust fractions 
represent the range of possible values derived from 
the SED fitting process (see \S\ref{methodology}).
\label{figmethodology}}
\end{center}
\end{figure}

Figure~\ref{figmethodology}a shows an example of our spectral
decomposition for the A8V target HD~142666; the two-component fit is
quite acceptable.  From the fitting results, only two parameters are
used in subsequent analysis: the K-band IR excess and the star
luminosity (in cases where good literature estimates are unavailable;
see Table~\ref{targets}).  In particular, the best-fit dust
temperature and flux are not directly used in visibility fitting since
the ring model is purely {\em geometric}.  We note that scattered
stellar light can not be distinguished from direct stellar emission in
the SED, thus the point source component might be somewhat
underestimated; this effect is minimized by observing at K band, where
scattering is much less efficient than for J or H bands.

Because YSOs are often variable sources, we were concerned with making
reliable estimates of the K-band IR excess based on non-contemporaneous
photometry.  We have adopted the
following procedure to conservatively estimate our observational
uncertainties.  We created multiple synthetic SEDs using various
combinations of visible (typically, B,V,R) and infrared photometric
datasets drawn from the literature.  For each SED we obtained an
estimate of the fraction of light at K-band arising from the
circumstellar dust through model-fitting.  The fitting results for the
IR excess naturally depended on the exact photometry data used, and
this variation was quantified (typically $\sim$10\%) and reported in
Table~\ref{keckresults2} for all targets.  The ``best estimate'' was
based on the SED with the most recent IR photometry (usually from
2MASS).

Once the NIR excess has been separately estimated, a ring model can be
fit to the KI visibility data with only one free parameter, the ring
diameter (having already adopted a 20\% ring thickness).  This process
is illustrated for HD~142666 in Figure~\ref{figmethodology}b, showing
fitting results for 3~different estimates of the fraction of K-band
emission coming from the dust component (``dust fraction'').
Table~\ref{keckresults2} contains the complete fitting results 
for all sources, listing the ring diameters along with errors
including both the visibility measurement error and our uncertainty in
dust fraction.  The reported uncertainty in the ring diameter is often
dominated by our uncertainty in estimating the dust fraction, instead
of $V^2$ measurement errors.

With our current single-baseline observations, we are unable to detect
disk elongations if present, such as those reported by
\citet{eisner2003,eisner2004}.  Based on these workers' data, we can
expect up to a 50\% variation in ring diameter depending on the
orientation of the disk on the sky, and this source of scatter will be
discussed further in \S\ref{sld}.

\section{Results}

The results of our model fits can be found in
Table~\ref{keckresults2}.  All but one source were resolved by the
Keck Interferometer (the transition object HD~141569 was unresolved), and the ring diameters
ranged from $\sim$1.5~mas to 4~mas.  In order to compare the ring
diameters of different sources, we have converted angular
diameters (milliarcseconds) into physical sizes (AU) using the 
estimated distances (drawn from the literature), and these
values are also tabulated.

A few of our survey targets (MWC~758, HD 141569, v1685 Cyg, v1977~Cyg)
were observed recently by \citet{eisner2004} using the Palomar Testbed
Interferometer (PTI).  Comparing average ring diameter results, we
find that PTI and KI results agree at the 1-2~$\sigma$
level. This reasonable agreement suggests that neither experiment is
contaminated with large-scale scattered light, since the PTI
experiment has a 20$\times$ larger field-of-view than the KI
observations.  While agreement between these two datasets is generally
satisfactory, we note an apparent inconsistency between the PTI and
KI data for v1685~Cyg along one particular position angle. 
The current PTI data has a relatively low signal-to-noise ratio, 
making a definitive comparison uncertain; future work will investigate this 
apparent discrepancy in more detail.

\subsection{Size-Luminosity Diagram}
\label{sld}

\begin{figure}[thb]
\begin{center}
%\epsscale{.5}
\includegraphics[angle=90,height=4in]{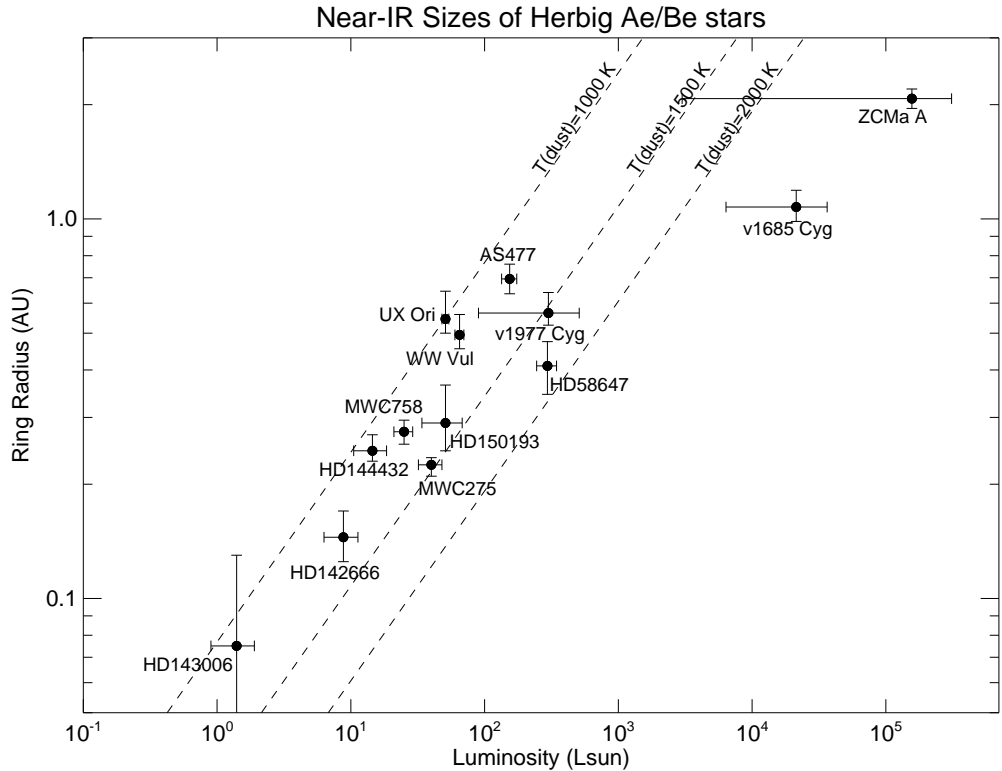}
%\includegraphics[angle=90,height=3in]{Figures/rleo2.epsi}
%      \plotfiddle{Figure/fig:astrometry.eps,rot=90}^M
%\plotfiddle{Figures/fig:astrometry.eps}{12cm}{1}{1}{0}{0}{-90}^M
\caption{\scriptsize 
Near-infrared sizes of Herbig Ae/Be stars.
The Herbig Ae/Be stars observed by the Keck Interferometer have been
located on this ``size-luminosity diagram.'' 
The plot symbols show the radius of dust
emission for the ``ring'' model discussed in the text.  The dashed lines
represent the expected inner edge of a dust disk truncated by dust
sublimation at temperatures $T_s=$1000\,K, 1500\,K, and 2000\,K,
assuming an optically-thin inner cavity and grey dust.
\label{figszlum}}
\end{center}
\end{figure}

In order to investigate specific accretion disk models, we wish to compare
the observed physical sizes of the NIR emission to model predictions.   Since 
the NIR emission should be dominated by hot dust
at the inner edge of the dusty disk, we expect the truncation radius to be
some function of the stellar luminosity.  
Thus,  we have plotted
our results on a size-luminosity diagram (Figure~\ref{figszlum})
using the parameters compiled in Table~\ref{targets}, 
as first described in \citet{monnier2002a}.
Since we have used a ring model for our emission geometry, the
ring radius (which we measured by fitting to the visibility data)
can be identified with the dust destruction radius,
the location where the dust temperature exceeds the 
sublimation temperature.

\begin{figure}[thb]
\begin{center}
%\epsscale{.5}
\includegraphics[angle=0,height=4in]{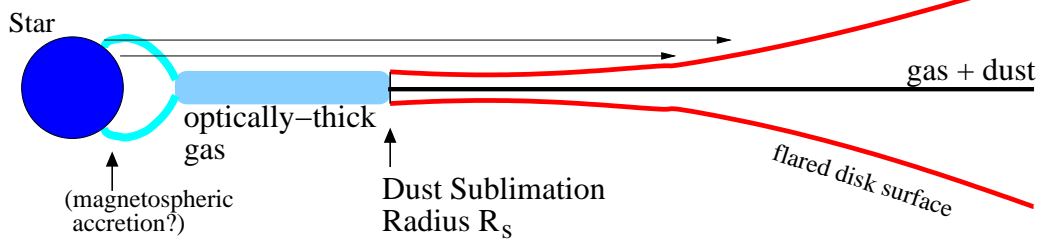}
%\includegraphics[angle=90,height=3in]{Figures/rleo2.epsi}
%      \plotfiddle{Figure/fig:astrometry.eps,rot=90}^M
%\plotfiddle{Figures/fig:astrometry.eps}{12cm}{1}{1}{0}{0}{-90}^M
\caption{\scriptsize 
Near-infrared sizes of Herbig Ae/Be stars compared to simple models.
Here we present schematics of the two disk models under consideration
in this paper.  {\em (top panel)} a. This panel shows a cross-section of
the inner disk region for the ``optically-thin cavity'' model discussed
in the text.  {\em (bottom panel)} b. Here, the cross-section of
the ``classical'' accretion disk model is shown.  This model is
nearly identical to the first, except the presence of optically-thick
gas in the midplane partially shields the innermost dust from
stellar radiation,
causing the dust sublimation radius ($R_s$) to shrink
for the same sublimation temperature ($T_s$).  
In order to place our results in a broader context, we have labelled
some additional relevant disk features (magnetospheric accretion columns, 
disk flaring, puffed-up
inner wall) which are not directly constrained by our study.
\label{disks}}
\end{center}
\end{figure}

In Figure~\ref{figszlum}, we compare our observational results to
predictions of a simple physical model for the dusty circumstellar
environment.  Here, we assume the star is surrounded by an
optically-thin cavity and that dust is distributed in a disk geometry
at larger radii.  The size of
the inner cavity is set by the dust sublimation radii $R_s$ (for
sublimation temperature $T_s$) and can be calculated from basic
radiation transfer (see Figure~\ref{disks}a for schematic drawing).
In this model, spherical dust grains at the inner edge are in
thermal equilibrium with the unobscured central star of luminosity
$L_\ast$, leading to a standard result \citep{monnier2002a}:

\begin{equation}
R_s = \frac{1}{2} \sqrt{\epsilon_Q} {\left( \frac{T_\ast}{T_s} \right) }^2 R_\ast
=1.1 \, \sqrt{\epsilon_Q} {\left(\frac{L_\ast}{1000\,L_\odot}\right) }^\frac{1}{2}
{\left(\frac{T_s}{1500\,K}\right)}^{-2}~AU
\label{eq1}
\end{equation}
where $\epsilon_Q = Q_{\rm{abs}} (T_\ast) / Q_{\rm{abs}} (T_s)$ 
the ratio of the dust absorption
efficiencies $Q(T)$ for radiation at color temperature T of the
incident and reemitted field respectively.

We note that the sublimation radius derived by \citet[][see Eq.
14]{dullemond2001} is a factor of 2 times larger than Eq. \ref{eq1},
since these workers assume the dust forms an optically-thick ``hot
inner wall'' at $R_s$.  Recent work by \citet{whitney2004} found
sublimation radii close to that expected in this optically-thick
limit.  In reality, $R_s$ will be between these limits depending how
abruptly the dust becomes optically thick as a function of radius.

Inspection of Figure~\ref{figszlum} reveals that Herbig targets between
$1\,\lsun<L_\ast<10^3\,\lsun$ have ring radii consistent with the
calculated dust sublimation radii $R_s$
for sublimation temperatures $T_s\sim$1000-1500K, and the
observed sizes are tightly correlated with stellar luminosity.
The data points in the Keck Interferometer size-luminosity diagram show 
much less
scatter around the $R\propto L^{\frac{1}{2}}$ trendlines 
than was seen in previous studies
\citep[especially,][]{monnier2002a}. We owe the reduced scatter to
the homogeneous nature of our dataset and the
improvements in the experimental methodology, including better
target vetting for reliable spectral types, higher angular resolution,
use of K band (instead of H band) and smaller field-of-view to reject
scattered light, and lastly higher signal-to-noise ratio.
These high-quality data offer definitive confirmation of the 
size-luminosity relations $R\propto L^{\frac{1}{2}}$ for Herbig Ae and late Be disks,
strong evidence for the optically-thin cavity model.

The remaining size scatter of about 50\% is likely due to inclination
effects, consistent with the range of elongations observed by
\cite{eisner2003,eisner2004}.  Further evidence for this comes from
the fact that UX~Ori, which is widely believed to be viewed at high
inclination angle \citep[e.g.,][]{natta1999}, is one of the most
extreme targets in Figure~\ref{figszlum}, showing the lowest apparent
sublimation temperature.  We note that our visible-light sensitivity limit
($R\simle 10.5$) introduces a bias against edge-on sources if the disk
heavily obscures the star.

We appreciate that the true NIR brightness distribution of YSO disks
may not be adequately described by the ring model adopted here,
despite our sound scientific motivations.  Only future imaging work by
IOTA, CHARA, and/or VLTI interferometers will unambiguously establish the
true emission geometry.
However, regardless of the exact emission morphology, the
size-luminosity trends presented here should still remain valid if the
targets in our sample share a common emission geometry.

\subsection{``Under-sized'' disks around high luminosity sources}

While the ``optically-thin cavity'' model can explain the observed
sizes of Herbig Ae and late Be stars, the higher luminosity (high-L)
sources clearly deviate from the model predictions.  These measured
disk sizes are many times too small to be consistent with the
size-luminosity relations found for the lower luminosity sources.
While our sample only contains two such sources, v1685~Cyg (B3) and
Z~CMa A (B0?), the results must be taken seriously given the
unambiguous discrepancy that can not be explained by known sources of
uncertainty (however, see \S\ref{zcma} for specific discussion of the
problematic Z~CMa system).

In order to investigate this further, we have calculated the dust
sublimation radii for an alternate disk model: the ``classical''
optically-thick, geometrically thin disk model for $T_s\sim$ 1500\,K and 1000\,K
\citep[e.g.,][]{hillenbrand1992,rmg2001}.  The main difference from the
previous model is that a thin disk of gas acts to shield the dusty
disk from direct stellar illumination. Thus, the expected
sublimation radii $R_s$ are significantly smaller than for the
corresponding optically-thin cavity model.  Figure~\ref{disks}
contains a sketch contrasting the two model geometries under
investigation here.

\begin{figure}[thb]
\begin{center}
%\epsscale{.5}
\includegraphics[angle=90,height=4in]{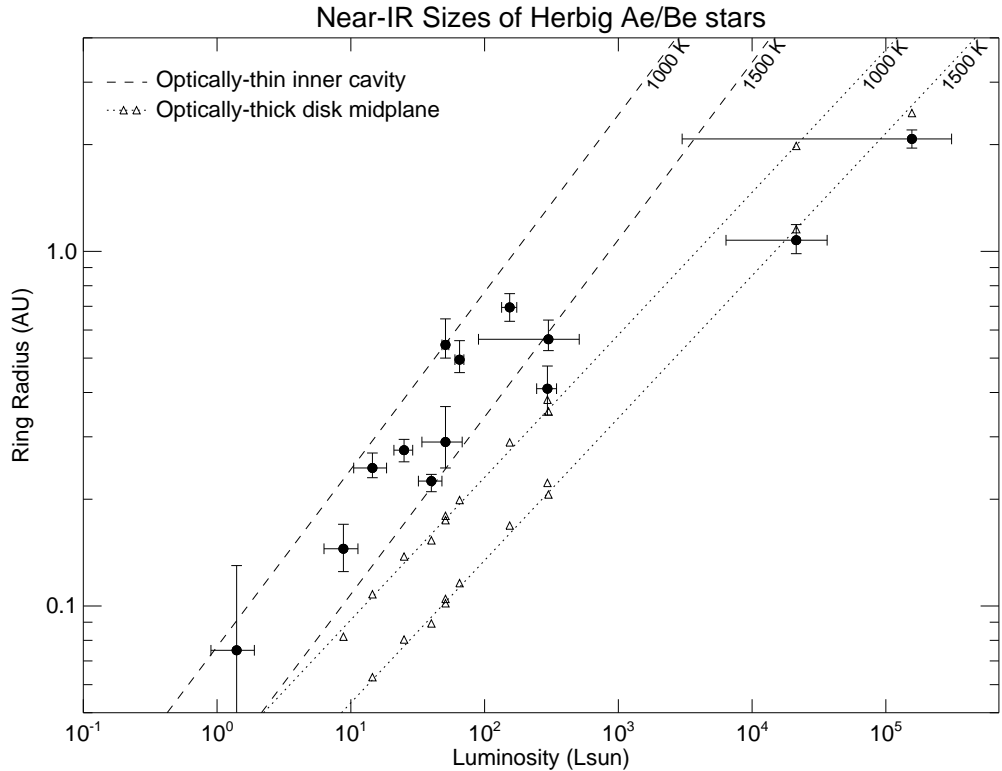}
%\includegraphics[angle=90,height=3in]{Figures/rleo2.epsi}
%      \plotfiddle{Figure/fig:astrometry.eps,rot=90}^M
%\plotfiddle{Figures/fig:astrometry.eps}{12cm}{1}{1}{0}{0}{-90}^M
\caption{\scriptsize The size-luminosity diagram here allows one to
compare the observed disk sizes (solid circles) with the predictions
of two simple disk models (see Figure~\ref{disks}).  
The dashed lines show the disk sizes
assuming an optically-thin inner disk (same as previous
Figure~\ref{figszlum}), while the dotted lines show mean relations for
the inner dust radius predicted for an optically-thick,
geometrically-thin inner disk model (open triangles show calculation
for individual sources).
Both models assume grey dust and results are shown for
dust sublimation temperatures $T_s=$1500\,K \& 1000\,K.
Note that if the inner wall of the ``optically-thin cavity'' model
is truly opaque 
as proposed by \citet{dullemond2001} then the inferred sublimation
temperatures are larger by $\sim$500~K (see \S\ref{sld} for more discussion).
\label{figszlum3}}
\end{center}
\end{figure}

The dust destruction radius for the classical accretion disk model is
not a pure function of luminosity (see above references for derivation
of analytical formulation) and thus a separate model estimate must be
made for each target based on specific stellar parameters.  The
results of this calculation are plotted in Figure~\ref{figszlum3}
along with the Keck Interferometer disk sizes and the
previously-derived size-luminosity relations (e.g., Eq. \ref{eq1}).
Here we see that the high-L targets are fit better by ``classical''
accretion disk models than by the optically-thin cavity models,
confirming recent analysis of \citet[][based on v1685~Cyg, MWC~1080,
and MWC~297]{eisner2004}.

We can look for further confirmation of this trend by re-considering
the results of \citet{monnier2002a}.  These authors also found many
high-L disks to have ``undersized'' emission, although a few notable targets
showed {\em order-of-magnitude} larger sizes than their high-L peers.
Notably, MWC~349 and LkH$\alpha$~101 (targets resolved by aperture
masking) were much larger than their counterparts measured with
long-baseline interferometry.  Perhaps these sources represent more
evolved systems where strong stellar winds have either cleared the
inner disk of gas or photo-evaporation has eroded the inner disk.
Regardless, our Keck Interferometer results reveal a population of
high-luminosity YSOs with disk sizes much smaller than possible for
models with optically-thin inner cavities.

In summary, our new Keck Interferometer data support the conclusions
of previous studies that some high-luminosity (early Herbig Be) stars
show evidence for significant gaseous inner disks.  There appears to
be a diversity of such disks, from those consistent with {\em
completely} optically-thick disk midplanes \citep[][and this
work]{eisner2004} to those with intermediate-tau inner cavities 
\citep[perhaps optically-thick only in the
ultraviolet; see][]{monnier2002a}.  We note that dust destruction radii
data alone can not directly constrain the geometry of the inner gaseous
material -- a completely optically-thick midplane has a similar effect
to an intermediate-tau spherical gas distribution.  In order to
distinguish between these scenarios, one must incorporate other
observations that also probe the inner accretion disk, such as {\em
mid-infrared} disk sizes \citep[e.g., recently reported
by][]{hinz2001,leinert2004} and high-resolution spectroscopy of the
molecular gas \citep[e.g.,][]{najita2003,brittain2003}.

\subsection{Comments on individual sources}

{\bf HD~141569:} HD~141569, the only unresolved target in our
survey, is thought to be in  a transition stage between pre-main
sequence disk and debris disk.  Although disk structures can be seen
in scattered light \citep{clampin2003}, SED modeling by \citet{li2003}
(and others) suggest that nearly all the NIR emission is
from the star and that the disk emits significant radiation only at
longer wavelengths. Our own SED-fitting and Keck interferometry data
support this picture.

{\bf UX~Ori:} UX~Ori is the prototype of pre-main sequence objects showing deep visual
minima interpreted as obscuration by dust clouds.  UXOR behavior is
thought to arise for YSOs (typically Herbig Ae/Be stars) which are
viewed at high inclination \citep[e.g.,][]{natta1999}, 
such that our line-of-sight partially
intercepts the accretion disk.  Under these conditions, we 
expect the simple ``ring'' emission
geometry (assumed here for visibility fitting) to break down.
As we discussed in \S\ref{sld}, UX~Ori has a relatively
large disk size, showing a deviation from the mean size-luminosity
relations of Figure~\ref{figszlum}.  The NIR
emission of UX~Ori is likely not ring-like and we may be seeing
scattered light if the inner disk emission is partially obscured by
the outer flared disk. Clearly, UX~Ori is a prime target for
interferometric imaging with VLTI and CHARA.

{\bf HD~58647:} According to Figure~\ref{figszlum}, this disk is
unusually small considering its luminosity.  In fact, it has the
hottest inferred dust sublimation temperature of the low-luminosity
Herbigs.  Recently, \citet{manoj2002} argued that this source is more
likely a classical Be star rather than a pre-main-sequence object.
Indeed, if the NIR emission is partly arising from free-free
(gas) emission, we would expect the observed interferometric size to
be small.

\label{zcma}
{\bf Z~CMa:} This binary source consists of a Herbig (Z~CMa A) and an
FU~Orionis object (Z~CMa B) in a close ($\sim 100$~milliarcsecond)
orbit \citep[e.g.,][]{garcia1999,koresko1991}.  
The Keck Interferometer was able to observe both sources
independently using the 50~milliarcsecond resolution of the adaptive
optics system (the FUOR component will be treated in 
Millan-Gabet et al., in preparation).
Determining the photometric contributions of the two
components separately is problematic due to strong variability and the
small angular separation. 

This source is also highly embedded
\citep{hartmann1989,whitney1993} and thus the applicability of the ``ring
model'' for describing the NIR emission is suspect.  In addition, the
spectral type and luminosity of the Herbig component is highly
uncertain -- with a factor of 100 in luminosity separating two
reasonable estimates.
We adopt a lower limit of 3000$\lsun$ here based on the bolometric
luminosity of the system \citep{hartmann1989}, while an upper limit of
310000$\lsun$ comes from de-reddening an assumed B0III spectral type
\citep{vda2004}.  Obviously this large luminosity uncertainty makes
definitive analysis of the Keck Interferometer data impossible, and we
encourage follow-up spectroscopic and
interferometric observations to confirm the properties of this unique and
challenging source.

\section{Conclusions}

We have definitively measured the near-infrared size-luminosity
relations for disks around Herbig Ae/Be stars for $L_\ast <
10^3~\lsun$.  Valid over more than 2 decades in stellar luminosity
$L$, the NIR sizes  obey the simple scaling relation $R\propto
L^{\frac{1}{2}}$.  This relation is predicted by the ``optically-thin
cavity'' model for YSO disks
\citep{tuthill2001a,natta2001,monnier2002a} and our results imply dust
sublimation temperatures in the expected range of $T_s \sim
1000-1500\,K$.

In contrast, the infrared sizes of circumstellar disks for the
high-luminosity sources in our sample ($L_\ast > 10^3~\lsun$) are more
consistent with {\em optically-thick} inner disks, supporting recent
conclusions of \citet{eisner2004}.  Significant gas in the inner disk
midplane could explain these observational results, although the gas
spatial distribution is not {\em directly} constrained here.
Exceptions to this trend are notable
(LkH$\alpha$~101 and MWC 349, reported elsewhere), perhaps signaling the clearing of the
inner gaseous disk by the strong stellar winds and ionizing radiation
from evolved O and early B stars.

Future work will progress on two fronts, one theoretical and the other
observational. We will focus on using state-of-the-art physical models
to fit SEDs and visibilities at both near-IR and mid-IR wavelengths.  Such
work will allow us to move beyond merely-qualitative tests of new disk
models \citep[e.g., the DDN models,][]{dullemond2001}, for instance by
quantifying the optical depth and geometry of the inner gaseous disk.
These studies will provide the density and temperature profiles needed
for studies of planet formation.

To advance observationally, more single-baseline data are still needed
for classical T~Tauri disks (Akeson et al., in preparation),
FU~Orionis stars (Millan-Gabet et al., in preparation), and also for
the highest luminosity sources.  In contrast, studies of Herbig Ae
disks will not be significantly advanced with more single-baseline
data, given the (now) large body of existing results; Herbig Ae disks
should now be {\em imaged} to make additional progress.  With closure
phase imaging from IOTA, VLTI and/or CHARA interferometers, we will test
predictions of
the next-generation of {\em physical} models, such as those incorporating a 
``puffed-up inner wall'' (DDN).

\acknowledgements

The authors wish to thank the all members of the Keck Interferometer
development team (JPL, MSC, WMKO) whose dedicated efforts made this
``shared-risk'' commissioning science possible.  This material is
based upon work supported by NASA under JPL Contracts 1236050 \&
1248252 issued through the Office of Space Science.  Data presented
herein were obtained at the W.M. Keck Observatory from telescope time
allocated to the National Aeronautics and Space Administration through
the agency's scientific partnership with the California Institute of
Technology and the University of California. The Observatory was made
possible by the generous financial support of the W.M. Keck
Foundation.  This research has made use of the SIMBAD database,
operated at CDS, Strasbourg, France. This publication makes use of
data products from the Two Micron All Sky Survey (2MASS), which is a joint
project of the University of Massachusetts and the Infrared Processing
and Analysis Center/California Institute of Technology, funded by the
National Aeronautics and Space Administration and the National Science
Foundation.  This work has made use of services produced by the
Michelson Science Center at the California Institute of Technology.
The authors wish to recognize and acknowledge the very significant
cultural role and reverence that the summit of Mauna Kea has always
had within the indigenous Hawaiian community.  We are most fortunate
to have the opportunity to conduct observations from this mountain.

\bibliographystyle{apj}
\bibliography{apj-jour,iKeck,KeckIOTA,Thesis,Review,Review2,HerbigSizes}
\clearpage

%\hspace{-.5in}
\begin{deluxetable}{lllllllll}
%\rotate
\tabletypesize{\scriptsize}
\tabletypesize{\tiny}
\tablecaption{Basic Properties of Targets\label{targets}}
%\tablewidth{0pt}
\tablehead{
\colhead{Source} & \colhead{RA (J2000)} & \colhead{Dec (J2000)} &
\colhead{V} & \colhead{K} &
\colhead{Spectral} & \colhead{Distance} &\colhead{Adopted} & \colhead{Photometry}\\
\colhead{Names} & & &\colhead{mag\tablenotemark{a}} & \colhead{mag\tablenotemark{a}} & \colhead{Type}  & \colhead{(pc)} & \colhead{Luminosity (L$_\odot$)} & \colhead{References} 
}
%\rotate
%\tabletypesize \scriptsize
%\rotate
\startdata
UX~Ori, HIP 23602 & 05 04 29.9908 &$-$03 47 14.280 & 9.6 & 7.2 & A3 (1) & 460 (10) & 51$\pm$3 (1) & 10, 15, 16, 17, 18, 19, 20, 21, 22  \\
MWC~758, HD~36112 & 05 30 27.5296 &$+$25 19 57.083 & 8.3 & 5.8 & A8V (9) & 200$^{+60}_{-40}$ (4,5) & 25$\pm$4 (14) & 15, 16, 19, 20, 21, 23\\
Z~CMa~A\tablenotemark{b},  HIP 34042  &  07 03 43.1619 & $-$11 33 06.209 & 9.9 & 3.8 & B0III (13) & 1050 (13) & 3000 -- 310000 (13,26) &  13, 20, 24, 25 \\
HD~58647, HIP 36068 &  07 25 56.0989  & $-$14 10 43.551 & 6.8 & 5.4 & B9IVe (3,4) & 280$^{+80}_{-50}$ (4,5) & 295$\pm$50 (14) &  15, 16, 19, 20, 21\\
HD~141569, HIP 77542 & 15 49 57.7489 & $-$03 55 16.360 & 7.0 & 6.8 & A0Vev (3,6) & 99$^{+9}_{-8}$ (4,5) & 18.5$\pm$1.0 (14) & 15, 16, 18, 19, 20, 21\\
HD~142666, v1026~Sco & 15 56 40.023 & $-$22 01 40.01& 8.8 & 6.1 & A8Ve (3,6,7) & 116 (7) & 8.8$\pm$2.5 (14) & 15, 16, 18, 19, 20, 21 \\ 
HD~143006, HIP 78244 & 15 58 33.4177 &  $-$09 00 12.174 & 8.4 & 7.1 & G5V (6) & 94$\pm$35 (5) & 1.4$\pm$0.5 (14) & 15, 16, 18, 19, 20, 21, 22 \\
HD~144432, HIP 78943 & 16 06 57.9575 & $-$27 43 09.806 & 8.2 & 5.9 & A9IVev (3) & 145 (8)  &  14.5$\pm$4.0 & 15, 16, 18, 19, 20, 21, 22\\
HD~150193, MWC 863A &  16 40 17.9221 &$-$23 53 45.180 & 8.9 & 5.5 & A2IVe (3) & 150$^{+50}_{-30}$ (4,5) &  51$\pm$17 (14) & 10, 15, 16, 18, 19, 20, 21, 22 \\
MWC~275, HD~163296 & 17 56 21.2879 & $-$21 57 21.880 & 6.9 & 4.8 & A1Vepv (3) & 122$^{+17}_{-13}$ (4,5) & 40$\pm$8 (14) &  10, 15, 16, 17, 18, 19, 20, 21, 22\\
WW~Vul, HD 344361 & 19 25 58.750 &$+$21 12 31.28 & 10.5 & 7.3 & A2IVe (3) & 550 (12) &  65$\pm$5 (1) &  15, 16, 17, 18, 19, 20, 21\\
v1685~Cyg, MWC~340 & 20 20 28.2473 &  $+$41 21 51.586 & 10.7 & 5.8 & B3 (1) & 980 (4) & 21400$\pm$15000 (1) & 10, 15, 16, 17, 18, 19, 20, 21\\
v1977~Cyg, AS~442 & 20 47 37.47 &$+$43 47 24.9 & 10.9 & 6.6 & B8V (3) & 700 (11) & 300$\pm$210 (14) & 15, 16, 17, 19, 20, 21  \\
AS~477, v1578~Cyg & 21 52 34.0993&  $+$47 13 43.612 & 10.2 & 7.2 & A0 (1) & 900 (2) & 154$\pm$20 (1)& 10, 15, 16, 17, 18, 19, 20, 21\\ 
\enddata 
\tablenotetext{a}{Many of the targets are variable stars and
these magnitudes (V band from Simbad, and K band from 2MASS) are merely
representative.}  
\tablenotetext{b}{Z CMa is a binary system \citep{millangabet2002} consisting of an Herbig Component (A) and an FU Orionis star (B).
The V and K magnitudes in the table are for the entire system, although our data reduction
isolate the contribution from the Herbig component.  The spectral type and luminosity
for Z~CMa A is highly uncertain -- see \S\ref{zcma} for discussion.}
\tablecomments{References: (1) \citet{hernandez2004}, (2) \citet{lada1985}, (3) \citet{mora2001}, (4) \citet{vda1998}, (5) \citet{hipparcos}, (6) \citet{dunkin1997a}, (7) \citet{meeus2001}, (8) \citet{perez2004}, (9) \citet{beskrovnaya1999}, (10) \citet{hillenbrand1992}, (11) \citet{terranegra1994}, (12) \citet{friedemann1993}, (13) \citet{vda2004}, (14) SED fitting, this work (15) \citet{morrison2001}, (16) \citet{kharchenko2001}, (17) \citet{morel1978}, (18) \citet{gezari1999}, (19) Simbad Astronomical Database, (20) 2MASS; \citet{cutri2003},
(21) Tycho-2; \citet{tycho2}, (22) DENIS Database, 2nd Release, (23) \citet{iras}, (24) \citet{millangabet2002}, (25) \citet{koresko1991}, (26) \citet{hartmann1989}
}
\end{deluxetable}
\clearpage

%% Text for table notes should follow after the \enddata but before
%% the \end{deluxetable}. Make sure there is at least one \tablenotemark
%% in the table for each \tablenotetext.

%\tablenotetext{a}{Sample footnote for table~\ref{tbl-1} that was generated
%with the deluxetable environment}
%\tablenotetext{b}{Another sample footnote for table~\ref{tbl-1}}%
%
%\tablecomments{Occasionally, authors wish to append a short
%paragraph of explanatory notes that pertain to the entire table, but
%which are different than the caption.  Such notes should be placed in
%a {\tt tablecomments} command like this.}
%

%\clearpage
\begin{deluxetable}{llllcc}

%\tabletypesize{\scriptsize}
\tablecaption{Calibrator Information\label{calibrators}
}
\tablewidth{0pt}
\tablehead{
\colhead{Source} &
\colhead{Calibrator} & \colhead{Spectral} & \colhead{V} & \colhead{K} &\colhead{Adopted Uniform Disk\tablenotemark{a}} \\
\colhead{Name} & 
\colhead{Name} & \colhead{Type} & \colhead{mag} & \colhead{mag} & \colhead{Diameter (mas)}    }
\startdata
UX~Ori & HDC33278 & G9V & 8.6 & 6.8 & 0.12$\pm$0.2 \\
 & HDC36003 & K5V & 7.7 & 4.8 & 0.55$\pm$0.1 \\
 & HDC26794 & K3V & 8.8 & 6.3 & 0.24$\pm$0.1 \\
MWC~758 & HDC27777 & B8V & 5.7 & 5.9 & 0.21$\pm$0.1 \\
 & HDC29645 & G0V & 6.0 & 4.6 & 0.41$\pm$0.2 \\
Z~CMa~A & HDC48286 & F7V & 7.0 & 5.7 & 0.32$\pm$0.1 \\
 & HDC52919 & K5V & 8.4 & 5.5 & 0.39$\pm$0.1 \\
 & HDC60491 & K2V & 8.2 & 5.9 & 0.30$\pm$0.1 \\
HD~58647 & HDC58461 & F3V & 5.8 & 4.9 & 0.39$\pm$0.1 \\
 & HDC62952 & F2V & 5.0 & 4.2 & 0.39$\pm$0.5 \\
HD~141569 \& HD~144432 & HDC139909 & B9.5V & 6.9 & 7.0 & 0.16$\pm$0.1 \\
 & HDC147550 & B9V & 6.2 & 6.3 & 0.21$\pm$0.1 \\
HD~142666 \& HD~150193 & HDC144641 & G3V & 8.0 & 6.5 & 0.15$\pm$0.1 \\
 & HDC134967 & A2V & 6.1 & 6.0 & 0.28$\pm$0.1 \\
HD~143006 & HD141107 & F2V & 7.7 & 6.9 & 0.17$\pm$0.1 \\
 & HD149149 & G6V & 8.6 & 7.0 & 0.12$\pm$0.1 \\
%HD~144432 & HDC139909 & B9.5V & 6.9 & 7.0 & 0.16$\pm$0.1 \\
% & HDC147550 & B9V & 6.2 & 6.3 & 0.21$\pm$0.1 \\
%HD~150193 & HDC144641 & G3V & 8.0 & 6.5 & 0.15$\pm$0.1 \\
% & HDC134967 & A2V & 6.1 & 6.0 & 0.28$\pm$0.1 \\
MWC~275  & HDC157546 & B8V & 6.3 & 6.5 & 0.17$\pm$0.1 \\
 & HDC174596 & A3V & 6.6 & 6.5 & 0.21$\pm$0.1 \\
WW~Vul & HDC181047 & G8V & 8.3 & 6.5 & 0.19$\pm$0.1 \\
 & HDC184198 & F7V & 8.2 & 6.9 & 0.14$\pm$0.1 \\
v1685~Cyg \& v1977~Cyg   & HDC199178 & G2V & 7.2 & 5.7 & 0.20$\pm$0.2 \\
v1685~Cyg & HIP102667 & K2V & 8.8 & 6.6 & 0.13$\pm$0.2 \\
 & HDC192985 & F5V & 5.9 & 4.8 & 0.38$\pm$0.1 \\
%v1977~Cyg & HDC199178 & G2V & 7.2 & 5.7 & 0.20$\pm$0.2 \\
v1977~Cyg & SAO50092 & K0V & 8.6 & 6.3 & 0.30$\pm$0.04 \\
 & HDC199998 & K2III & 8.4 & 5.7 & 0.40$\pm$0.1 \\
AS~477 & HDC201456 &   F8V & 7.9&  6.6 &0.18$\pm$ 0.1  \\
       & HIP109034 &   K4III& 9.5 & 6.2 &  0.35$\pm$0.5 \\
       & HDC199178 &   G2V  & 7.2 & 5.7 &  0.20$\pm$0.2\\
\enddata
\tablenotetext{a}{All diameter estimates were made using {\em getCal}, which is
maintained and distributed by the
Michelson Science Center (http://msc.caltech.edu).
The diameters were estimated by fitting the spectral energy distributions with
simple blackbody models.
}
\end{deluxetable}

%% If you use the table environment, please indicate horizontal rules using
%% Do not put multiple tabular environments within a single table.
%% The optional \label should appear inside the \caption command.

\begin{deluxetable}{llllc}
\tabletypesize{\scriptsize}
\tablecaption{Keck Interferometer Visibilities\label{keckresults}
}
\tablewidth{0pt}
\tablehead{
\colhead{Source} & \colhead{U.T. } &\multicolumn{2}{c}{Projected Baseline} & \colhead{Visibility-Squared} \\
\colhead{Name} & \colhead{Date}  &  \colhead{U (m)} & \colhead{V (m)} &  \colhead{$\lambda_0=2.18 \mu$m, $\Delta\lambda= 0.3 \mu$m} 
}
\startdata
UX~Ori    & 2004 Jan 07 & 55.547 & 64.015 & 0.483$\pm$0.052\\
          &             & 56.336 & 63.635 & 0.476$\pm$0.042\\
MWC~758   & 2002 Oct 24 & 50.790 & 67.981 & 0.360$\pm$0.088\\
          &             & 49.974 & 68.668 & 0.337$\pm$0.039\\
Z~CMa~A   & 2004 Apr 03 & 28.806 & 52.420 & 0.172$\pm$0.015\\
          & 2004 Apr 05 & 34.159 & 53.142 & 0.186$\pm$0.077\\
HD~58647  & 2004 Apr 02 & 30.502 & 49.851 & 0.639$\pm$0.098\\
          &             & 27.501 & 49.411 & 0.734$\pm$0.091\\
HD~141569 & 2003 Apr 17 & 55.224 & 62.474 & 0.978$\pm$0.065\\
          &             & 54.102 & 62.180 & 1.056$\pm$0.105\\
HD~142666 & 2004 Mar 05 & 56.538 & 58.909 & 0.562$\pm$0.059\\
          &             & 56.219 & 56.564 & 0.445$\pm$0.054\\
          &             & 55.784 & 55.363 & 0.586$\pm$0.092\\
          &             & 55.647 & 55.064 & 0.507$\pm$0.066\\
          &             & 53.392 & 51.838 & 0.604$\pm$0.033\\
HD~143006 & 2004 Apr 02 & 40.097 & 42.871 & 0.931$\pm$0.197\\
          &             & 36.677 & 41.635 & 0.863$\pm$0.200\\
HD~144432 & 2003 Apr 17 & 54.460 & 49.100 & 0.368$\pm$0.018\\
          &             & 54.167 & 48.628 & 0.380$\pm$0.016\\
HD~150193 & 2004 Mar 05 & 54.858 & 52.468 & 0.205$\pm$0.023\\
MWC~275   & 2003 Apr 17 & 55.592 & 55.001 & 0.188$\pm$0.010\\
          &             & 54.449 & 53.157 & 0.218$\pm$0.015\\
          &             & 53.189 & 51.683 & 0.218$\pm$0.013\\
          &             & 51.713 & 50.306 & 0.232$\pm$0.015\\
WW~Vul    & 2003 Aug 09 & 37.683 & 74.410 & 0.642$\pm$0.063\\
          &             & 33.914 & 75.528 & 0.558$\pm$0.059\\
v1685~Cyg & 2002 Jun 27 & 41.541 & 72.969 & 0.397$\pm$0.069\\
          &             & 33.288 & 77.825 & 0.420$\pm$0.108\\
v1977~Cyg & 2002 Oct 24 & 45.133 & 69.368 & 0.636$\pm$0.032\\
          &             & 34.998 & 76.533 & 0.640$\pm$0.041\\
AS~477    & 2002 Oct 24 & 40.923 & 71.723 & 0.703$\pm$0.025\\
          &             & 35.262 & 75.529 & 0.673$\pm$0.054
\enddata
%\tablenotetext{a}{The best-fit UD dia)
%}
\end{deluxetable}

%\clearpage
\begin{deluxetable}{llllll}

%\tabletypesize{\scriptsize}
\tablecaption{Herbig Ae/Be Disk Properties
\label{keckresults2}
}
\tablewidth{0pt}
\tablehead{
\colhead{Source} & \colhead{Dust Fraction} & \multicolumn{2}{c}{Ring Diameter}  \\
\colhead{Name} & \colhead{at K-band\tablenotemark{a}}  &  \colhead{mas} & \colhead{AU} 
}
\startdata
UX~Ori    & 0.70$^{+0.08}_{-0.16}$ & 2.36$^{+0.43}_{-0.20}$ & 1.09$^{+0.20}_{-0.09}$ \\
MWC~758   & 0.72$^{+0.02}_{-0.04}$ & 2.75$^{+0.22}_{-0.19}$ & 0.55$\pm$0.04 \\
Z~CMa~A   & 0.975$\pm$0.025        & 3.95$\pm$0.24 & 4.15$\pm$0.25 \\
HD~58647  & 0.54$^{+0.10}_{-0.04}$ & 2.93$\pm$0.45 & 0.82$\pm$0.13 \\ 
HD~141569 & 0.05$\pm$0.05          &  $<$20 & $<$2 \\
HD~142666 & 0.61$^{+0.15}_{-0.13}$ & 2.52$^{+0.41}_{-0.31}$ & 0.29$^{+0.05}_{-0.04}$ \\
HD~143006 & 0.53$^{+0.11}_{-0.03}$ &  1.63$^{+1.12}_{-1.00}$ & 0.15$^{+0.11}_{-0.09}$ \\
HD~144432 & 0.62$^{+0.02}_{-0.07}$ & 3.37$^{+0.32}_{-0.17}$ & 0.49$^{+0.05}_{-0.03}$ \\
HD~150193 & 0.67$^{+0.18}_{-0.17}$ & 3.84$^{+1.03}_{-0.58}$ & 0.58$^{+0.15}_{-0.09}$ \\
MWC~275   & 0.71$^{+0.07}_{-0.01}$ & 3.70$^{+0.14}_{-0.25}$ & 0.45$^{+0.02}_{-0.03}$ \\
WW~Vul    & 0.88$^{+0.03}_{-0.14}$ & 1.80$^{+0.24}_{-0.15}$ & 0.99$^{+0.13}_{-0.08}$ \\
v1685~Cyg & 0.94$^{+0.02}_{-0.09}$ & 2.19$^{+0.23}_{-0.18}$ & 2.15$^{+0.23}_{-0.18}$ \\
v1977~Cyg & 0.94$^{+0.02}_{-0.17}$ & 1.61$^{+0.22}_{-0.12}$ & 1.13$^{+0.15}_{-0.08}$ \\
AS~477    & 0.86$^{+0.03}_{-0.04}$ & 1.55$^{+0.14}_{-0.13}$ & 1.39$^{+0.13}_{-0.12}$ \\
\enddata
\tablenotetext{a}{Best estimate for fraction of K-band light coming from circumstellar material based
on most recent photometry.  Upper and lower limits are based on 
SED fitting to diverse data sets and represent the range of possible values given 
historical variability.}

\end{deluxetable}
%\begin{table}

%\begin{center}

%\caption{More terribly relevant tabular information.\label{tbl-2}}

%\begin{tabular}{crrrrrrrrrrr}

%\tableline\tableline

%Star & Height & $d_{x}$ & $d_{y}$ & $n$ & $\chi^2$ & $R_{maj}$ & $R_{min}$ &

%\multicolumn{1}{c}{$P$\tablenotemark{a}} & $P R_{maj}$ & $P R_{min}$ &

%\multicolumn{1}{c}{$\Theta$\tablenotemark{b}} \\

%\tableline

%1 &33472.5 &-0.1 &0.4  &53 &27.4 &2.065  &1.940 &3.900 &68.3 &116.2 &-27.639\\

%2 &27802.4 &-0.3 &-0.2 &60 &3.7  &1.628  &1.510 &2.156 &6.8  &7.5 &-26.764\\

%3 &29210.6 &0.9  &0.3  &60 &3.4  &1.622  &1.551 &2.159 &6.7  &7.3 &-40.272\\

%4 &32733.8 &-1.2\tablenotemark{c} &-0.5 &41 &54.8 &2.282  &2.156 &4.313 &117.4 &78.2 &-35.847\\

%5 & 9607.4 &-0.4 &-0.4 &60 &1.4  &1.669\tablenotemark{c}  &1.574 &2.343 &8.0  &8.9 &-33.417\\

%6 &31638.6 &1.6  &0.1  &39 &315.2 & 3.433 &3.075 &7.488 &92.1 &25.3 &-12.052\\

%\tableline

%\end{tabular}

%% Any table notes must follow the \end{tabular} command.

%

%\tablenotetext{a}{Sample footnote for table~\ref{tbl-2} that was

%generated with the \LaTeX\ table environment}

%\tablenotetext{b}{Yet another sample footnote for table~\ref{tbl-2}}

%\tablenotetext{c}{Another sample footnote for table~\ref{tbl-2}}

%\tablecomments{We can also attach a long-ish paragraph of explanatory

%material to a table. Use \tablerefs to append a list of references. (See

%the notes to the next table for an example.)}

%\end{center}

%\end{table}

\end{document}